\documentclass{nature}

\usepackage{amsmath}
\usepackage{amssymb}
\usepackage{gensymb}
\usepackage{graphicx}
\usepackage{upgreek}
\usepackage{lineno}
\usepackage{color}
\usepackage{longtable}
\usepackage{caption}
\def\mnras{Mon. Not. R. Astron. Soc.}
\def\apj{Astrophys. J.}

\def\apjs{Astrophys. J. Suppl.}
\def\aj{Astron. J.}
\def\araa{Annu. Rev. Astron. Astrophys.}
\def\aap{Astron. Astrophys.}

\def\pasp{Publ. Astron. Soc. Pacif.}
\def\nat{Nature}

\def\HI{H\,{\sc i}}
\def\HII{H\,{\sc ii}}

\def\MgII{Mg\,{\sc ii}}
\def\NaI{Na\,{\sc i}}
\def\NII{N\,{\sc ii}}
\def\OIII{O\,{\sc iii}}

\title{A population of highly energetic transient events in the centres of active galaxies}

\author{E. Kankare$^{1*}$, R.~Kotak$^{1}$, S.~Mattila$^{2,3}$, P.~Lundqvist$^{4}$, M.~J.~Ward$^{5}$, M.~Fraser$^{6,7}$, A.~Lawrence$^{8}$, S.~J.~Smartt$^{1}$, W.~P.~S.~Meikle$^{9}$, A.~Bruce$^{8}$, J.~Harmanen$^{2}$, S.~J.~Hutton$^{10}$, C.~Inserra$^{1,11}$, T.~Kangas$^{2}$, A.~Pastorello$^{12}$, T.~Reynolds$^{2,13}$, C.~Romero-Ca\~nizales$^{14,15}$, K.~W.~Smith$^{1}$, S.~Valenti$^{16}$, K.~C.~Chambers$^{17}$, K.~W.~Hodapp$^{17}$, M.~E.~Huber$^{17}$, N.~Kaiser$^{17}$, R.-P.~Kudritzki$^{17}$, E.~A.~Magnier$^{17}$, J.~L.~Tonry$^{17}$, R.~J.~Wainscoat$^{17}$ and C.~Waters$^{17}$}

\begin{document}

\maketitle

\sloppy

\begin{affiliations}
 \item Astrophysics Research Centre, School of Mathematics and Physics, Queen's University Belfast, Belfast BT7 1NN, UK. 
 \item Tuorla Observatory, Department of Physics and Astronomy, University of Turku, V\"ais\"al\"antie 20, FI-21500 Piikki\"o, Finland. 
 \item Finnish Centre for Astronomy with ESO (FINCA), University of Turku, V{\"a}is{\"a}l{\"a}ntie 20, FI-21500 Piikki{\"o}, Finland. 
 \item Oskar Klein Centre, Department of Astronomy, AlbaNova, Stockholm University, SE-10691 Stockholm, Sweden. 
 \item Department of Physics, Durham University, South Road, Durham DH1 3LE, UK. 
 \item Institute of Astronomy, University of Cambridge, Madingley Road, Cambridge CB3 0HA, UK. 
 \item School of Physics, O'Brien Centre for Science North, University College Dublin, Belfield, Dublin 4, Ireland. 
 \item Institute for Astronomy, University of Edinburgh, Royal Observatory, Blackford Hill, Edinburgh EH9 3HJ, UK. 
 \item Imperial College of Science Technology and Medicine, London SW7 2AZ, UK. 
 \item Department of Physics and Astronomy, University College London, Gower Street, London WC1E 6BT, UK. 
 \item Department of Physics and Astronomy, University of Southampton, Southampton, Hampshire, SO17 1BJ, UK. 
 \item INAF $-$ Osservatorio Astronomico di Padova, Vicolo dell'Osservatorio 5, I-35122 Padua, Italy. 
 \item Nordic Optical Telescope, Apartado 474, E-38700 Santa Cruz de La Palma, Spain. 
 \item Millennium Institute of Astrophysics, 7500011 Santiago, Chile. 
 \item N\'ucleo de Astronom\'{\i}a de la Facultad de Ingenier\'{\i}a y Ciencias, Universidad Diego Portales, Avenida Ej\'ercito 441, Santiago, Chile. 
 \item Department of Physics, University of California, Davis, CA 95616, USA. 
 \item Institute for Astronomy, University of Hawaii at Manoa, Honolulu, HI 96822, USA. 

 $^{*}$email: e.kankare@qub.ac.uk
\end{affiliations}

\begin{abstract}
Recent all-sky surveys have led to the discovery of new types of transients. These include stars disrupted by the central supermassive black hole, and supernovae that are 10-100 times more energetic than typical ones. However, the nature of even more energetic transients that apparently occur in the innermost regions of their host galaxies is hotly debated\cite{drake11,dong16,leloudas16}. Here we report the discovery of the most energetic of these to date: PS1-10adi, with a total radiated energy of $\sim 2.3\times10^{52}$~erg. The slow evolution of its light curve and persistently narrow spectral lines over $\sim$3 yr are inconsistent with known types of recurring black hole variability. The observed properties imply powering by shock interaction between expanding material and large quantities of surrounding dense matter. Plausible sources of this expanding material are a star that has been tidally disrupted by the central black hole, or a supernova. Both could satisfy the energy budget. For the former, we would be forced to invoke a new and hitherto unseen variant of a tidally disrupted star, while a supernova origin relies principally on environmental effects resulting from its nuclear location. Remarkably, we also discovered that PS1-10adi is not an isolated case. We therefore surmise that this new population of transients has previously been overlooked due to incorrect association with underlying central black hole activity.
\end{abstract}


We discovered the optical transient PS1-10adi at right ascension 20 h 42 min 44.74 s and declination 15$\degree$ 30$'$ 32.1$''$ (equinox J2000.0) in the Panoramic Survey Telescope and Rapid Response System 1 (Pan-STARRS1, or PS1) 3$\pi$ Faint Galaxy Supernova Survey\cite{inserra13} on 2010 August 15. Within the root mean square errors, PS1-10adi appeared to be coincident with the centre of Sloan Digital Sky Survey (SDSS) J204244.74+153032.1, and was $\gtrsim$2 mag brighter at peak in all optical bands than the SDSS data release 7 (DR7)\cite{abazajian09} reference magnitudes. Over a period of $\sim$1,000 d, its light curves evolve slowly and smoothly (Fig.~\ref{fig:abs}, Supplementary Fig. 1); after this phase the quiescent host galaxy starts to dominate the brightness. The early spectra are dominated by a blue continuum and narrow (900 km s$^{-1}$) Balmer lines, consistent with a redshift $z = 0.203 \pm 0.001$. The shape of the Balmer lines at early epochs is indicative of broadening by electron scattering\cite{dessart15}. Around $\sim$200~d, the Balmer line profiles show a pronounced asymmetry in the red wing (Fig.~\ref{fig:halpha}). Spectra taken at epochs $\gtrsim$3 yr are dominated by the quiescent host, and display signatures of both active galactic nuclei (AGN) and star-forming galaxies (see Methods). Compared with this quiescent level, the Balmer line and \MgII\ $\lambda\lambda$2796,2803 line fluxes increased, without any corresponding change in the [\OIII] $\lambda$5007 or other forbidden lines (Supplementary Fig. 2).

The U- to M-band photometry of PS1-10adi are well described by two blackbody components (see Methods). Over $\sim$1,000 d of monitoring, the hot and warm components evolved from $\sim$11,000 K to $\sim$8,000 K, and $\sim$2,500 K to $\sim$1,200 K, respectively. The blackbody radius of the hot and warm components peaked at $\sim$8$ \times 10^{15}$ cm and $\sim$1.3$ \times 10^{17}$ cm, respectively. The total blackbody luminosity (Fig.~\ref{fig:bol}) declined slowly and exponentially. Integrating over the blackbody evolution yields a radiated energy of $1.7 \times 10^{52}$~erg; however, this is an underestimate given the early ultraviolet (UV) excess ($\sim$20\%), and a missing rise-time contribution ($\sim$10\%) (see Methods and Supplementary Fig. 3). Thus, we infer the total radiated energy to be $2.3 \pm 0.5 \times 10^{52}$~erg. 

Spectroscopically, PS1-10adi bears similarities both to narrow-line Seyfert 1 galaxies\cite{osterbrock85} and to certain types of supernovae (type IIn)\cite{schlegel90} that show signatures of ejecta interacting with dense surroundings. \MgII\ emission (Supplementary Fig. 2) is commonly observed in both AGN\cite{ulrich97} and type IIn supernovae\cite{fransson14} and is therefore not a discriminant. 

We first explore whether PS1-10adi could be linked to reprocessed emission from either an AGN or a tidal disruption event (TDE)\cite{rees88}. We begin by considering AGN variability. AGN typically vary by only a few tenths of a magnitude, and do so stochastically\cite{macleod12} (Fig.~\ref{fig:comp}). This cannot account for the smooth $\gtrsim$2 mag brightening of PS1-10adi. The increase in the continuum brightness of `changing look quasars' arises from changes in the accretion rate or line-of-sight extinction\cite{macleod16}, and is directly and positively correlated with the emission from the AGN broad-line region. For PS1-10adi, this is not the case (Fig.~\ref{fig:halpha}): the H$\alpha$ emission at peak is as narrow (full-width at half maximum (FWHM) $\sim$ 900 km s$^{-1}$) as the underlying narrow-line region (FWHM$_{[\mathrm{O III}]~\lambda 5007}$ $\sim$ 700 km s$^{-1}$), and significantly narrower than the broad-line region component ($\sim$2,800 km s$^{-1}$) as estimated from the quiescent host galaxy spectrum at +1,600 d. The narrow-line region of Seyfert galaxies lies at $\sim10^{2}-10^{3}$ pc (ref.~\cite{netzer90}), corresponding to a light travel time of $>10^{2}$ yr, and cannot therefore vary on the observed timescales of PS1-10adi. Note that the emission lines in the broad-line region are expected to become somewhat narrower (FWHM $\propto$ $L^{-1/4}$), where the underlying AGN luminosity, $L$, increases due to gas at higher distances being photoionized\cite{denney09}. However, the observed luminosity increases by approximately a factor of $\sim$10, which would lead to $\sim$1,600 km s$^{-1}$ emission lines from the broad-line region of the PS1-10adi host, that we do not see. To obtain features that are even narrower, by approximately a factor of $\sim$2 (as observed), would require a luminosity increase by a factor of $\gtrsim$100. Additionally, such a luminosity variation is not thought to be accompanied by the appearance of a kinematically distinct component that we do observe. Furthermore, high-state AGN show a UV excess associated with thermal emission of the accretion disk close to $\sim$1,500~\AA, and have a distinctive break in the continuum slope near 5,000~\AA\ (ref.~\cite{vandenberk01}). No such break is seen in the spectra of PS1-10adi at any epoch, and our UV observations suggest that the spectral energy distribution (SED) peaks at $\sim$2,500~\AA. We conclude that known classes of AGN variability cannot account for PS1-10adi. If this is a new type of AGN behaviour, then the underlying physical mechanism remains to be identified, as do the relevant timescales.

Gravitational tidal forces close to a central black hole can shred a star that passes too close to it. Such events are expected to be bright as they are powered by the accretion luminosity of a fraction of the stripped stellar material onto a massive black hole\cite{phinney89}. Compared with PS1-10adi, canonical TDE light curves decline too rapidly (Fig.~\ref{fig:bol}), but it is conceivable that some TDEs could have slower decline rates. However, the blackbody temperatures of PS1-10adi are lower, and evolve faster than expected for TDEs. More importantly, the optical spectra of TDE candidates are either relatively featureless, or show broad (a FWHM of thousands of km s$^{-1}$) H or He emission lines\cite{gezari12,arcavi14} as expected for material close to a black hole. Reported TDEs have been found primarily in poststarburst galaxies\cite{arcavi14}, and do not display signatures indicative of dense ambient material. Even the transient ASASSN-15lh\cite{dong16,leloudas16} which has comparable peak brightness, has markedly different spectra and a distinctively different host galaxy.

Based on a phenomenogical reprocessing model\cite{loeb97} (see Methods), we estimate the amount of material required to reprocess radiation from a power source (AGN or TDE). To reproduce the peak luminosity, a shell or layer of $\sim$9 solar masses ($M_{\odot}$), and an accretion rate of $\sim0.2$ $M_{\odot}$ yr$^{-1}$ is required. However, the blackbody radius of PS1-10adi at peak is roughly consistent with the size of a Seyfert broad-line region\cite{netzer90}, $10^{-2} - 10^{-1}$ pc, which would give rise to broad emission lines (a FWHM of thousands of km s$^{-1}$), inconsistent with the observations (900 km s$^{-1}$). Furthermore, the source of the red shoulder of the Balmer lines is not clear, and disfavours a purely reprocessing shell without interaction as the source of emission for PS1-10adi.

We now discuss whether the observed characteristics of PS1-10adi could result from the interaction of expanding ejecta with a dense medium. In the classical interaction model, the ejecta crash into dense surrounding material, resulting in an outward-moving forward shock, and a reverse shock that recedes into the ejecta. This results in the efficient conversion of kinetic energy into radiative output\cite{chevalier94}, with the narrow emission lines arising from the slow-moving ambient medium. We interpret the appearance of a red shoulder (Fig.~\ref{fig:halpha}) in the Balmer lines as the shock in the expanding ejecta. The absence of such a prominent broad feature in the weak blue wing is suggestive of a non-spherical ejecta geometry, rather than being due to electron scattering. The red shoulder cannot be attributed to [\NII] $\lambda$6584 given the absence of the [\NII] $\lambda$6548 line and the similarity of the Balmer line profiles. The asymmetric nature of PS1-10adi does not allow us to model the data reliably, leading to degeneracies in model parameters. However, the inferred luminosity of PS1-10adi suggests a large mass of surrounding material. A number of different mechanisms can contribute to the build-up of material around a massive star, some of which are discussed below.

It has been proposed\cite{portegies07,vandenheuvel13} that runaway mergers of massive stars in dense and young stellar clusters can give rise to very massive H-rich circumstellar medium, resulting in ultrabright supernovae. Suitable dynamical conditions are expected in massive clusters that reside in the high-density central regions of an active galaxy. An exciting possibility is that a dense AGN environment can provide suitable conditions for high interstellar-medium pressure as well as the photoionizing radiation to trap a large fraction of the mass lost from the progenitor\cite{mackey14}. Furthermore, the interstellar-medium density of the narrow-line region of Seyfert galaxies\cite{netzer90} can itself be comparable (electron density $n_{\mathrm{e}} = 10^6$ electrons cm$^{-3}$) to the circumstellar medium around interacting supernovae, thereby providing material to slow the ejecta.

Interestingly, it may also be possible to recreate a type-IIn-like scenario via a TDE. Approximately 50 \% of the stellar material falling towards a black hole in a TDE is initially loosely bound to the black hole. If the radiative cooling of the bound material is inefficient, only a small fraction ($\ll$1) can accrete onto the black hole\cite{metzger16}. High-energy radiation arising from the accretion powers an outflow of the initially bound material with a velocity of $\sim$10$^{4}$ km s$^{-1}$; that is, the shell of material becomes unbound and mimics expanding ejecta. The model is qualitatively similar to an engine-powered supernova. In a Seyfert galaxy, the broad-line region gas clouds have high densities ($n_{\mathrm{e}} \gtrsim 10^8$ electrons cm$^{-3}$)\cite{netzer90}. If swept up by expanding TDE material, the shock interaction can give rise to narrow emission lines and slow light-curve evolution, as observed in many narrow-line supernovae. Passive elliptical galaxies lack this high-density environment. Thus, such events should preferentially occur in the high-pressure environments of active host galaxies. Indeed, even for intrinsic black hole variability, provided a suitable -- but as yet unknown --  mechanism is found for outflowing material, interaction with the broad-line region clouds could result in narrow-line emission spectra. We conclude that, for both the supernova and TDE cases, interaction with a surrounding medium can account for the observed properties, although some unexplained points remain, especially for the TDE case.

We searched through the PS1 and other transient databases and discovered a population of transients with characteristics that are remarkably similar to those of PS1-10adi (see Supplementary Table 1 and Methods for details of four further candidates). So far, these have mostly been ignored and attributed to normal AGN activity. We propose that they are a distinct, and probably not uncommon, class of transients that have not been recognised as such until now (Fig.~\ref{fig:comp}). PS1-10adi appears to be the most extreme case when compared with transients such as  CSS100217:102913+​404220 (hereafter, CSS100217) and PS1-13jw. The former is also spectroscopically similar to PS1-10adi\cite{drake11}. Intriguingly, all three host galaxies display AGN characteristics and line ratios that suggest ongoing star formation (Supplementary Fig. 4). The Large Synoptic Survey Telescope will discover hundreds of PS1-10adi-like transients, and the exquisite resolution of the Extremely Large Telescope (0.006$''$ in the J band) will allow the nuclear regions of $z=0.1$ galaxies to be probed at the 10 pc scale, opening up systematic studies of this new population of nuclear transients.\\

\begin{addendum}
 \item[Additional Information] \textbf{Correspondence and requests for materials} should be addressed to E.K.
 \item We thank G. Ferland, B. M\"{u}ller, K. Nilsson, M.-\'A. P\'erez-Torres and K. Poppenhaeger for discussions. E.K. and R.K. acknowledge support from the Science and Technology
Facilities Council (STFC; ST/L000709/1). M.F. acknowledges the support of a Royal Society -- Science Foundation Ireland University Research Fellowship. This work was partly supported by the European Union FP7 programme through the European Research Council (ERC) grant number 320360. S.J.S acknowledges ERC grant 291222 and STFC grants ST/I001123/1 and ST/L000709/1. J.H. acknowledges financial support from the Finnish Cultural Foundation. C.R-C. acknowledges support by the Ministry of Economy, Development and Tourism's Millennium Science Initiative through grant IC120009, awarded to The Millennium Institute of Astrophysics, Chile, and from the Comisi\'on Nacional de Investigaci\'on Cient\'ifica y Tecnol\'ogica through the Fondo Nacional de Desarrollo Cient\'ifico y Tecnol\'ogico grant 3150238. The PS1 surveys were made possible through contributions of the Institute for Astronomy, the University of Hawaii, the Pan-STARRS Project Office, the Max Planck Society and its participating institutes the Max Planck Institute for Astronomy, Heidelberg, and the Max Planck Institute for Extraterrestrial Physics, Garching, The Johns Hopkins University, Durham University, the University of Edinburgh, Queen's University Belfast, the Harvard-Smithsonian Center for Astrophysics, the Las Cumbres Observatory Global Telescope Network Inc., the National Central University of Taiwan, the Space Telescope Science Institute, NASA (National Aeronautics and Space Administration) under Grant No. NNX08AR22G issued through the Planetary Science Division of the NASA Science Mission Directorate, the National Science Foundation under Grant No. AST-1238877, the University of Maryland, Eotvos Lorand University and the Los Alamos National Laboratory. The Catalina Sky Survey is funded by NASA under Grant No. NNG05GF22G issued through the Science Mission Directorate Near-Earth Objects Observations Program. The Catalina Real-Time Transient Survey is supported by the US National Science Foundation under grants AST-0909182 and AST-1313422. This work is based on observations made with the NOT, the LT, the WHT, the WISE, Swift and the Karl G. Jansky Very Large Array.
 \item[Author Contributions] E.K. performed the data analysis and wrote the manuscript. R.K., S.M., P.L., M.J.W., M.F., A.L., S.J.S., W.P.S.M., S.J.H., C.I. and A.P. contributed to the physical interpretation. A.B. carried out follow-up observations with the WHT. C.R-C. carried out the radio data reductions. J.H., T.K. and T.R. carried out follow-up observations with the NOT. S.V., K.W.S. and S.J.S. built the Pan-STARRS Transient Science Server hosted at Queen's University Belfast. K.C.C., K.W.H., M.E.H., N.K., R.-P.K., E.A.M., J.L.T., R.J.W. and C.W. are PS1 builders. 
\end{addendum}

\clearpage

\renewcommand\figurename{\footnotesize \textbf{Figure}}
\renewcommand{\thefigure}{\textbf{\footnotesize \arabic{figure}}}

\begin{figure}
\centering
  \includegraphics[width=0.5\linewidth]{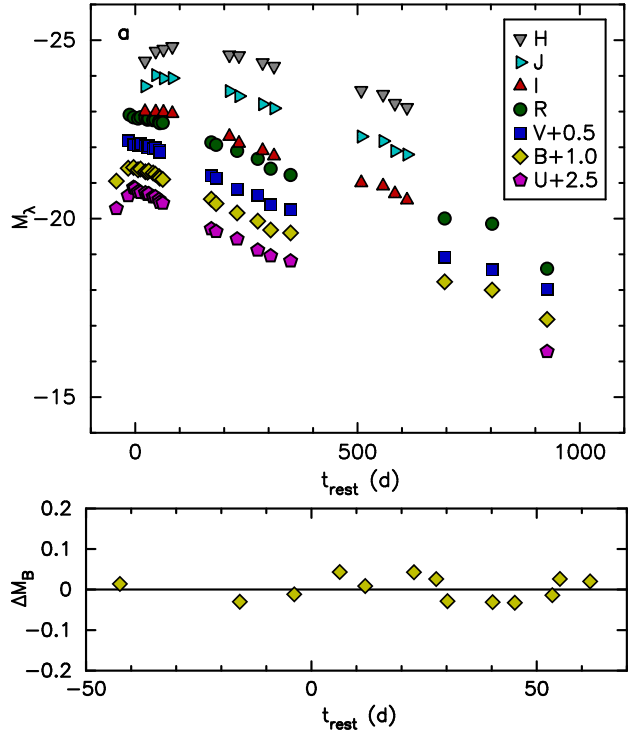}
  \includegraphics[width=0.5\linewidth]{fig1b.ps}
\caption{\textbf{Absolute magnitude $M_{\lambda}$ (Vega system) light curves of PS1-10adi.} \textbf{a}, The light curves are corrected for redshift, contribution of the host galaxy, Galactic extinction and luminosity distance. A K correction has also been applied. The rest-frame epoch, $t_{\mathrm{rest}}$, is given relative to the estimated time of optical maximum (see Methods). This yields peak optical magnitudes close to $-23$ mag. The NIR peak occurred roughly 100 d later, reaching an H-band magnitude of $M_{\mathrm{H}} \approx -25$ mag. The different symbols show different filters, and the UBV-band light curves are shifted for clarity as indicated in the key. \textbf{b}, Residuals $\Delta M_{\mathrm{B}}$\,from a low-order polynomial fit around the peak of the B-band light curve show that the evolution is smooth. Over a $\sim$3 yr monitoring period, no short-term fluctuations are evident. The 1$\sigma$ uncertainties are smaller than the symbols.}
 \label{fig:abs}
\end{figure}

\clearpage

\begin{figure}
\centering
 \includegraphics[width=0.5\linewidth]{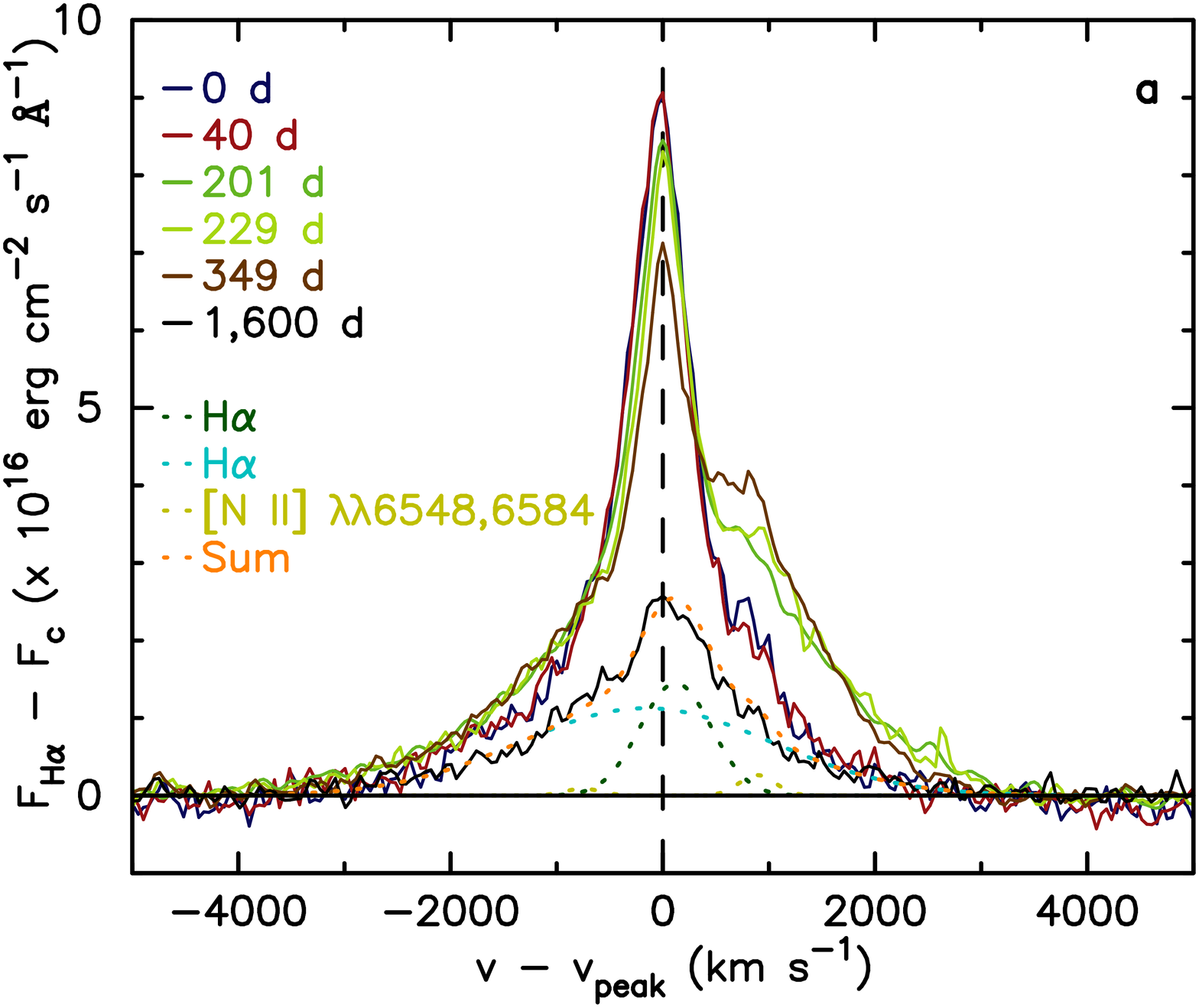}
 \includegraphics[width=0.5\linewidth]{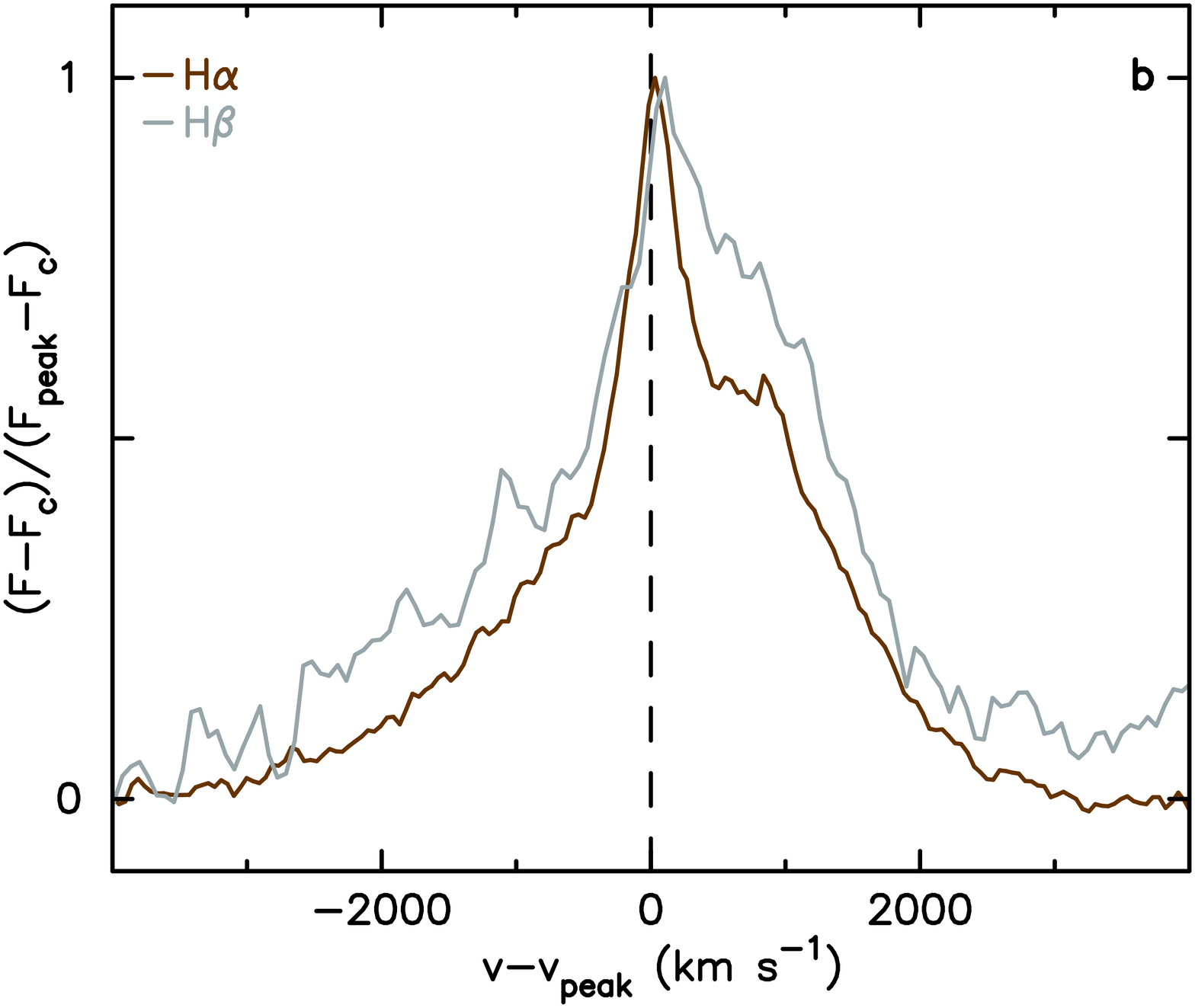}
  \caption{\textbf{Evolution of the continuum-subtracted H$\alpha$ line profile of PS1-10adi at selected epochs.} \textbf{a}, The Balmer lines at early epochs are narrow (900 km s$^{-1}$). The +201 to +349 d spectra display an increasingly prominent red shoulder in the H$\alpha$ profile that subsequently fades away. The fitted components and their sum of the quiescent host spectrum at +1,600 d are shown by the dotted curves, including a $\sim$2,800 km s$^{-1}$ broad-line region component. The continuum flux, $F_{\mathrm{c}}$, has been subtracted from the H$\alpha$ flux, $F_{\mathrm{H}\alpha}$. The velocity, $v$, is given relative to the line peak velocity, $v_{\mathrm{peak}}$. \textbf{b}, H$\alpha$ and H$\beta$ profiles, where $F$ is the flux, normalized with the peak flux, $F_{\mathrm{peak}}$, at +349 d, showing the similarity of the red shoulder of the Balmer line profiles.}
 \label{fig:halpha}
\end{figure}

\clearpage

\begin{figure}
\centering
 \includegraphics[width=0.5\linewidth]{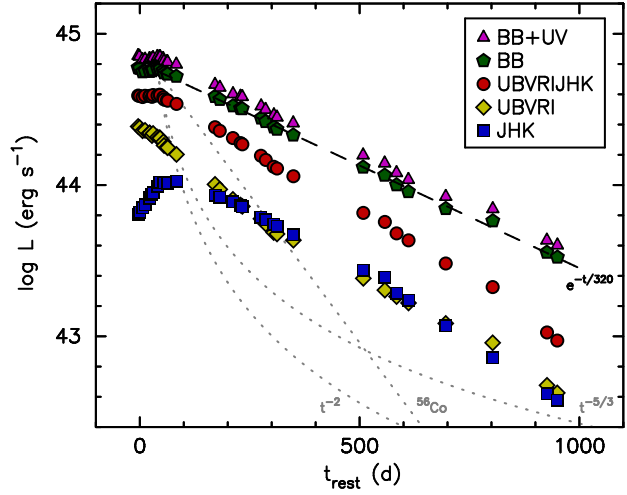}
 \caption{\textbf{Pseudobolometric JHK-, UBVRI- and UBVRIJHK-band and blackbody luminosity evolution of PS1-10adi.} The total hot and warm blackbody luminosity has a decline rate of $L_{\mathrm{BB}}(t) \propto e^{-t/320}$. The expected light curve evolution from a selection of scenarios is also shown: the canonical bolometric decline rate of tidal disruption flares\cite{phinney89} is $L \propto t^{-5/3}$; radioactive decay of $^{56}$Co $L \propto e^{-t/111.26}$ synthesized in supernova explosions; and superluminous supernova\cite{quimby11,gal-yam12} models of magnetar (a rapidly spinning and highly magnetized neutron star) spin-down energy\cite{kasen10} have a $L \propto t^{-2}$ decline rate. For PS1-10adi, the observed luminosity evolution is slower than that of any of the above. The blackbody luminosity with (BB+UV) and without (BB) an additional estimated 20\% UV component are also shown.}
 \label{fig:bol}
\end{figure}

\clearpage

\begin{figure}
  \includegraphics[width=0.99\linewidth]{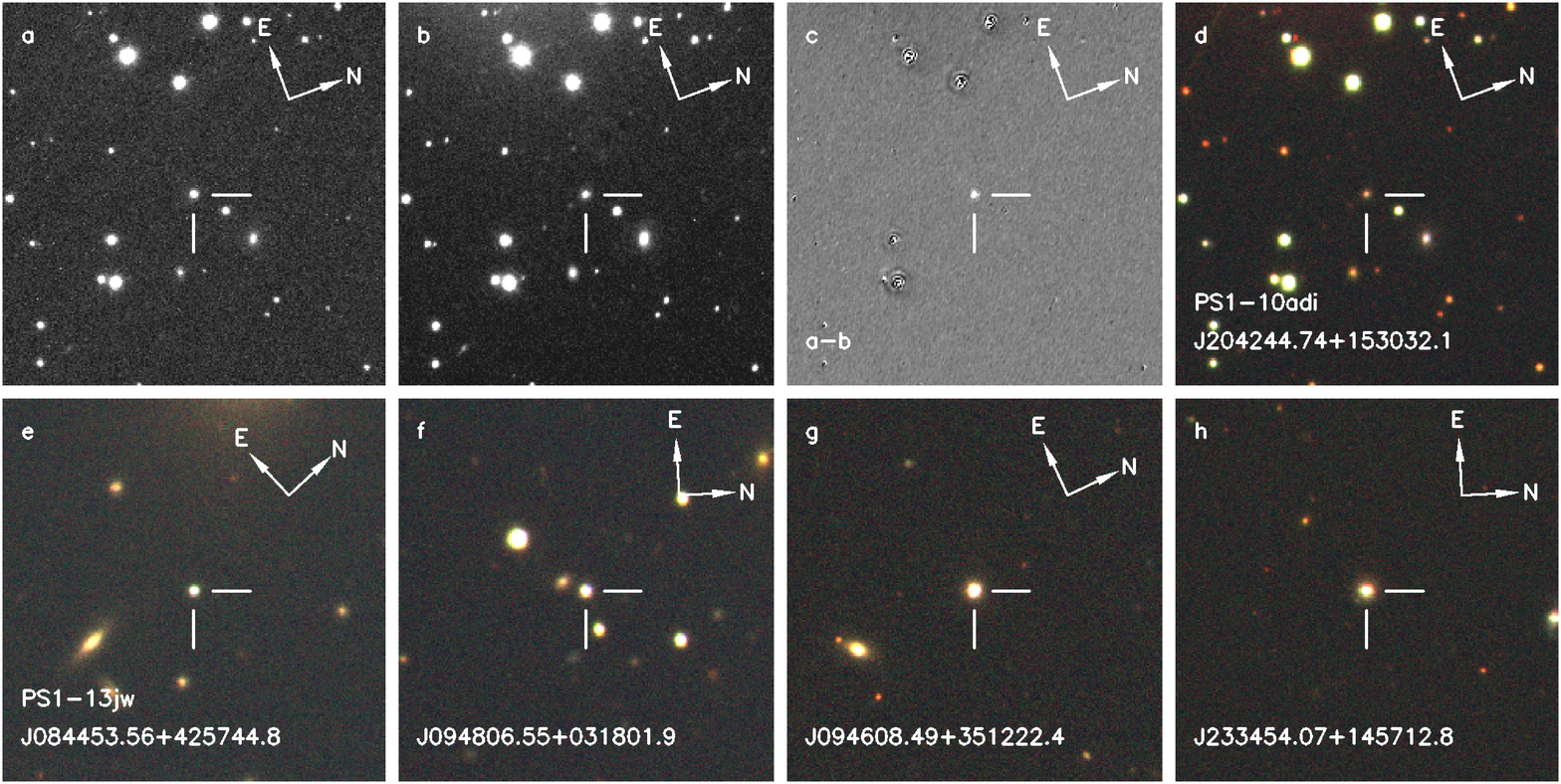}
  \includegraphics[width=0.49\linewidth]{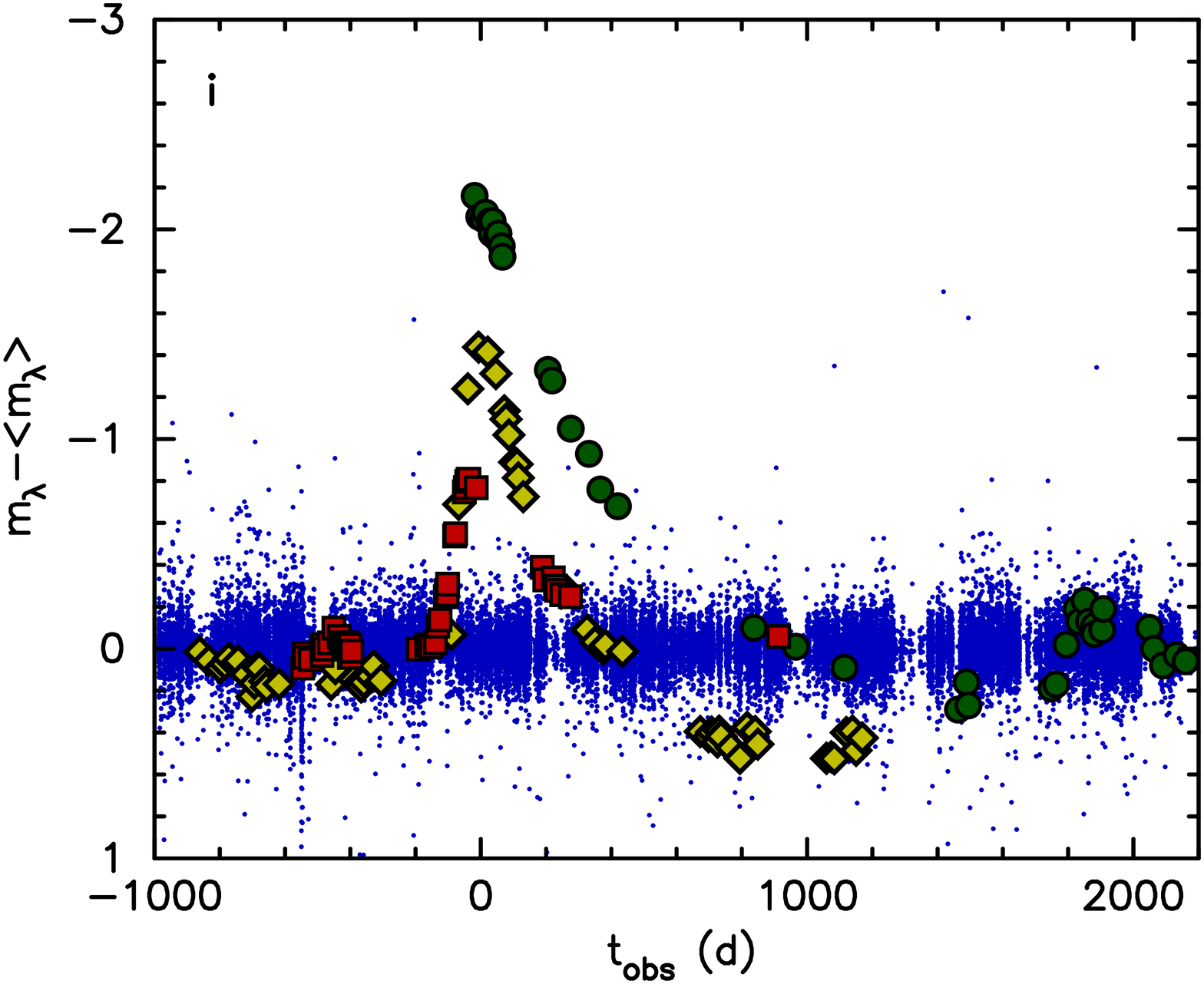}
  \includegraphics[width=0.49\linewidth]{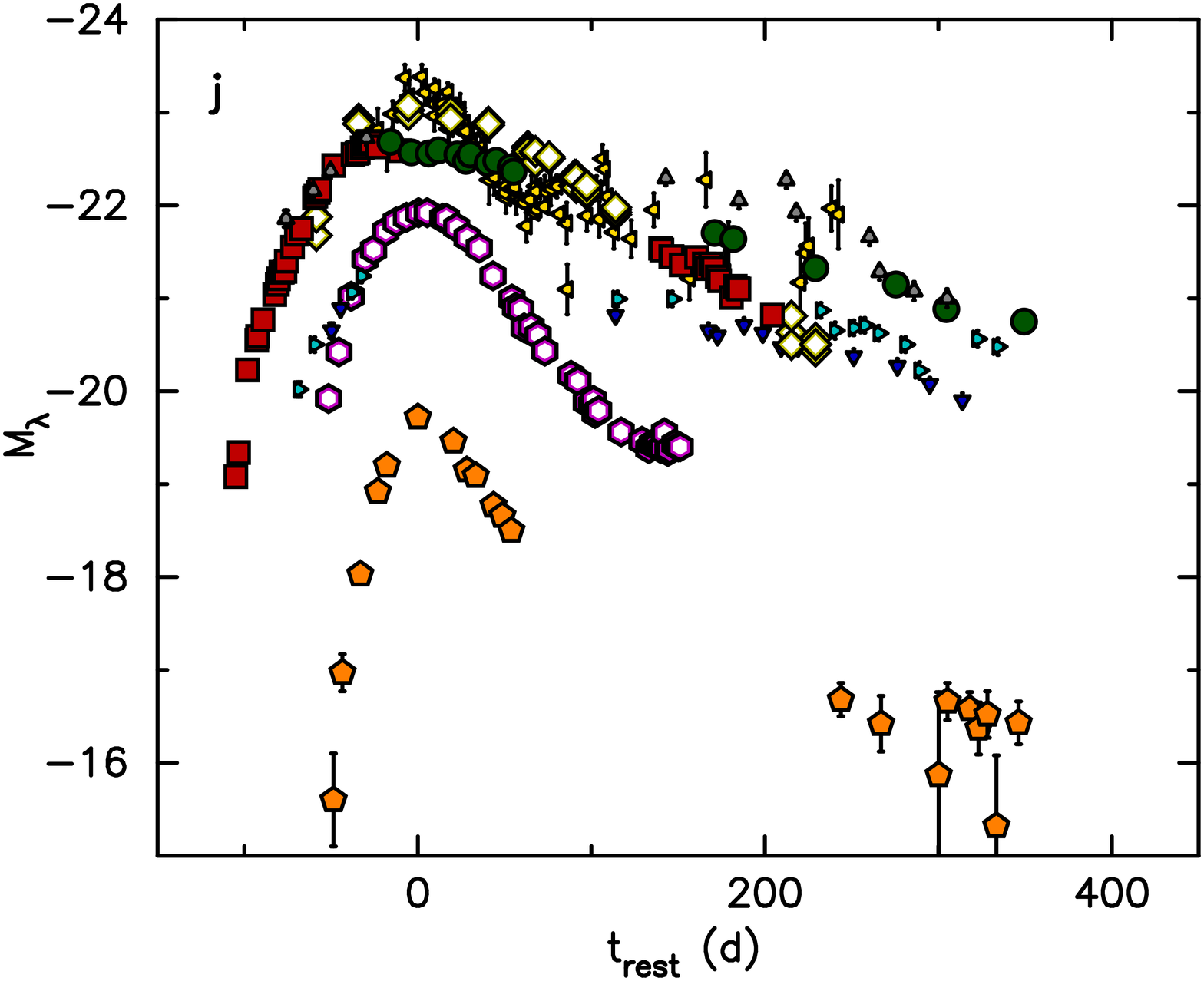}
\begin{minipage}{0.1\textwidth}
\hfill
\end{minipage}
\hfill
\begin{minipage}{0.25\textwidth}
\includegraphics[width=\linewidth]{fig4d.ps}
\end{minipage}
\hfill
\begin{minipage}{0.2\textwidth}
\hfill
\end{minipage}
\hfill
\begin{minipage}{0.20\textwidth}
\includegraphics[width=\linewidth]{fig4e.ps}
\end{minipage}
\hfill
\begin{minipage}{0.05\textwidth}
\hfill
\end{minipage}
\hfill
\end{figure}

\clearpage

\begin{figure}
 \caption{\textbf{Sample of PS1-10adi-like transients.} \textbf{a--b}, R-band images of the PS1-10adi field at two epochs: +182 d (\textbf{a}) and +1,216 d (\textbf{b}). \textbf{c}, Difference between the two epochs, clearly showing PS1-10adi. \textbf{d--h}, False-colour SDSS \textit{gri} images of the host galaxies (see Supplementary Table 1). All fields are 1.5$'$ $\times$ 1.5$'$. \textbf{i}, The normalized light curves of PS1-10adi, CSS100217 (ref.~\cite{drake11}) and PS1-13jw (see Methods), compared to 660 narrow-line Seyfert 1 galaxies ($<$18 mag) from the Catalina Surveys\cite{drake09} DR2 (blue dots). A representative error for narrow-line Seyfert 1 galaxies is shown in the key. The average quiescent magnitude, $\langle m_{\lambda} \rangle$, has been subtracted from the observed magnitudes, $m_{\lambda}$. The observer-frame epoch, $t_{\mathrm{obs}}$, is given relative to the estimated light-curve maximum. \textbf{j}, Absolute light-curve comparison including SN 2006gy\cite{smith07} (a superluminous type IIn supernova), PS1-10jh\cite{gezari12} (a TDE), ASASSN-15lh\cite{dong16,leloudas16} (a H-poor superluminous supernova or a TDE) and three further PS1-10adi-like candidates J094806, J094608 and J233454 (see Methods) from Catalina DR2. The light curves are in R-, \textit{r}-, V$_\mathrm{CSS}$- and V-bands as shown in the key. The 1$\sigma$ uncertainties are typically smaller than the symbols for named transients. For PS1-10adi and PS1-13jw, the errors computed in quadrature include contributions from point spread function and zeropoint variations.}
 \label{fig:comp}
\end{figure}

\clearpage 

\begin{methods}

\subsection{Basic parameters.} 
This study assumes a Hubble constant $H_{0}$ = 69.6 km s$^{-1}$ Mpc$^{-1}$, a cosmological constant $\Omega_{\Lambda}$ = 0.714 and a matter density $\Omega_{\rm M}$ = 0.286\cite{bennett14}. The redshift $z$ = 0.203 $\pm$ 0.001 from the Balmer lines of PS1-10adi corresponds to a luminosity distance $D_{L}$ = 1.00 $\pm$ 0.01 Gpc and a distance modulus of $\mu$ = 40.00 $\pm$ 0.02 mag. The redshift is in agreement with the SDSS DR7\cite{abazajian09} photometric $z$ = 0.219 $\pm$ 0.025 of the host galaxy SDSS J204244.74+153032.1. For the other transients in our sample (see Supplementary Table 1), we adopt the spectroscopic redshifts from SDSS. From the recent calibration of the Galactic dust map\cite{schlafly11}, Galactic line-of-sight extinctions are adopted, for example a V-band extinction $A_{V}$ = 0.281 mag for PS1-10adi. The spectra of PS1-10adi do not show a feature due to \NaI\ $\lambda\lambda$5890,5896 doublet, which is consistent with low host galaxy extinction. We therefore only apply a correction only for Galactic extinction, and likewise for the other transients. 

\subsection{Optical and near-infrared imaging.} 
Imaging observations were reduced in a standard manner. Nordic Optical Telescope (NOT) observations with ALFOSC were reduced using the quba pipeline\cite{valenti11}, and StanCam data were reduced using basic iraf tasks. Liverpool Telescope (LT) data products of the RATCam and IO:O instruments were automatically processed by the LT pipeline. LT imaging using the \textit{uBVri} filters was calibrated directly into the ultraviolet-blue-visual-red-infrared (UBVRI) system. Early imaging was obtained with the PS1 gigapixel camera (GPC1) as part of the 3$\pi$ survey in PS1 photometric system filters $g_{\mathrm{P1}}$, $r_{\mathrm{P1}}$ and $i_{\mathrm{P1}}$ (ref.~\cite{tonry12}). The \textit{gri} calibrated GPC1 magnitudes were converted into BVRI magnitudes using magnitude transformations\cite{jester05}. The near-infrared (NIR) data from the NOTCam instrument at the NOT were reduced with the external notcam package (version 2.5) within iraf. Point spread function photometry of the imaging data was carried out using the quba pipeline. We estimated the optical peak epoch, $t_{\mathrm{peak}}$, of PS1-10adi by fitting a third-order polynomial to the observed V-band light curve close to maximum, which yielded a rough $t_{\mathrm{peak}}$ estimate of Julian day 2455443. The optical and NIR photometry of PS1-10adi and its host galaxy are reported in Supplementary Tables 2 and 3.

By around day 1,200 in the observer frame, the evolution of PS1-10adi had plateaued at values consistent with the optical SDSS DR7 reference magnitudes. We consider the tail phase to be dominated by the (quiescent) host galaxy. Therefore, we adopt the observed magnitudes at +1,242~d (rest frame) as the host galaxy magnitudes, $m_{\mathrm{host}}$, and subtract this contribution from the observed magnitudes, $m_{\mathrm{obs}}$, in a standard way, to obtain the magnitudes, $m$, of the transient: $m = -2.5 \log_{10}(10^{-0.4 m_{\mathrm{obs}}} - 10^{-0.4 m_{\mathrm{host}}})$. 

\subsection{Space-based observations.} 
UV imaging data of PS1-10adi were obtained with the Swift Gamma-Ray Burst Mission space telescope using the UVOT instrument. The {\sc HEASoft} tasks were used to sum the individual exposures and carry out the aperture photometry. The resulting UV photometry is listed in Supplementary Table 4. Fortuitously, the field of PS1-10adi has been observed at 3.4~$\upmu$m and 4.6~$\upmu$m with the postcryogenic Wide-field Infrared Survey Explorer (WISE) space telescope. We obtained multi-epoch photometry from the ALLWISE and NEOWISE catalogues for PS1-10adi in individual WISE frames, and calculated the weighted average magnitudes in these bands for effective epochs of observations. 

\subsection{Bolometric evolution.}
For the pseudobolometric light curves, the observed magnitudes were corrected for Galactic extinction, converted to fluxes at effective wavelengths and corrected for redshift. Integrating the SED for each epoch yields the bolometric evolution. For epochs where no measurements were available, these were estimated primarily via linear interpolation. 

\subsection{Blackbody approximation.} 
To approximate the blackbody evolution of PS1-10adi, the quiescent host galaxy contribution was subtracted from the observed magnitudes, which were subsequently converted to fluxes following corrections for Galactic extinction and redshift. Two blackbody components were required to explain the SED (Supplementary Fig. 3), and were derived by minimizing the $\chi^{2}$ value of the fit for each epoch of photometry. During the first 1,000~d of evolution after the optical maximum, both hot and warm blackbody components decrease in temperature ($T_{\mathrm{hot}}$ and $T_{\mathrm{warm}}$, respectively), with $T_{\mathrm{hot}}$ evolving from $\sim$11,000~K to $\sim$8,000~K and $T_{\mathrm{warm}}$ from $\sim$2,500~K to $\sim$1,200~K. The radius of the hot component, $R_{\mathrm{hot}}$, decreases from $\sim$6$ \times 10^{15}$~cm to $\sim$1$ \times 10^{14}$~cm. The radius of the warm component, $R_{\mathrm{warm}}$, increases from $\sim$5$ \times 10^{16}$~cm to $\sim$1.3$ \times 10^{17}$~cm within $\sim$200~d and then stays fairly constant. A plausible explanation for this behaviour is that the warm component arises from reprocessed optical radiation by surrounding dust. The initial value of $T_{\mathrm{warm}} \approx $~2,500~K is high enough for some dust to be evaporating. With time, $R_{\mathrm{warm}}$ reaches a constant radius at $\sim$200~d coincident with $T_{\mathrm{warm}} \approx $~2,000~K, suggestive of the NIR emission being dominated by amorphous carbon grains, which have a high evaporation temperature ($\sim$2,000$-$3,000~K depending on the particle size), in contrast to that of silicate dust ($\sim$1,500~K). The peak luminosity of PS1-10adi suggests\cite{fransson14} a dust evaporation radius of $\sim$3 $\times 10^{17}$~cm, which is consistent with $R_{\mathrm{warm}}$ at +200~d. Similar NIR component has been observed, for example, in type IIn SN 2010jl with early $T_{\mathrm{warm}} \approx $̃2,300~K followed by a decline\cite{gall14}. Since the line-of-sight extinction of PS1-10adi appears to be low, the dust distribution can be asymmetric/clumpy.

The Swift UV points at +27 d and +80 d lie above the two-component blackbody fit. We estimate that this UV excess contributes an additional $\sim$20\% to the total luminosity. An exponential fit to the IR emission yields a rise time of $\sim$115 d (similar to the optical rise time of PS1-13jw) and suggests an additional contribution of $\sim$10\% to the total radiated energy of PS1-10adi.

\subsection{K correction.} 
Due to the lack of high cadence time series of optical and NIR spectra, the K corrections for the host subtracted light curves were derived iteratively from the observed photometry. Each epoch of photometry was fitted with a two-component blackbody as described above, to approximate the underlying SED from optical to NIR. These rest-frame blackbodies were redshifted back to the observer frame and were used to derive the K corrections. The \textit{BVRIJHK} filter magnitudes were corrected to \textit{UBVRIJH}. The observed U-band (and UV) magnitudes were not K corrected, since the shape of the SED in the UV region is uncertain. Likewise, the PS1-13jw \textit{griz} magnitudes were K corrected to \textit{ugri} and transformed\cite{jester05} to the Johnson-Cousins system for Fig.~\ref{fig:comp}. For CSS100217, the K corrections were taken from the literature\cite{drake11}, while PS1-10jh, ASASSN-15lh and other Catalina Real-Time Transient Survey\cite{drake09} targets were K corrected using $-2.5 \log_{10}(1+z)$.

\subsection{Optical spectroscopy.} 
We acquired a spectroscopic time series (Supplementary Fig. 2) of PS1-10adi using ALFOSC at the NOT, and the ISIS instrument at the William Herschel Telescope (WHT). The data reduction included bias subtraction and flat field correction. Wavelength calibration of the extracted one-dimensional spectra was carried out using arc lamp exposures; small offsets were derived to this calibration by cross-correlation with sky lines. Relative flux calibration was carried out using the sensitivity curves of the instrument derived from similarly processed observations of well-known spectroscopic standard stars, obtained with an identical setup. The above-mentioned steps were executed within the quba pipeline. Absolute flux calibration was performed by integrating over the spectra with the appropriate broadband filter transmission functions, and deriving scaling factors by comparing them with the observed photometry. Extinction and redshift corrections were carried out using standard iraf tasks. Line identifications are based on those of type IIn SN 2009kn (ref.~\cite{kankare12}).

\subsection{X-ray observations.} 
Our UVOT observations of PS1-10adi were accompanied by 0.3$-$10~keV observations with the X-ray Telescope onboard Swift. We downloaded the photon counting-mode Level 2 event data products from the Swift archive. For the four epochs of 2010 data during the PS1-10adi event, no signal was observed at the location of PS1-10adi. The 3$\sigma$ upper limits were estimated with the sosta task in the HEASoft package ximage, and likewise for the 2015 July 18 epoch. However, sosta found 2.6$\sigma$ and 1.4$\sigma$ sources close to the location of PS1-10adi, in the July 19 and 20 datasets, respectively. The possible detection on July 19 yields a source intensity of $3.5 \pm 1.4 \times 10^{-3}$~counts~s$^{-1}$ or $1.8 \pm 0.7 \times 10^{43}$~erg~s$^{-1}$ (full 0.3$-$10~keV range), corrected for poinst spread function, sampling dead time and vignetting. The X-ray sosta measurements are listed in Supplementary Table 4. The count rates were converted into flux using the HEASARC WebPIMMS tool, assuming a power law photon index $\Gamma = 2$, and adopting a Galactic \HI\ column density\cite{kalberla05} $N_{\mathrm{H}} = 6.48 \times 10^{20}$~cm$^{-2}$.

Optical and NIR observations of PS1-10adi around 1,800 d (observer frame) revealed an episode of slow brightening; we name this event as the `2015 bump'. This light-curve evolution is consistent with the optical variability of narrow-line Seyfert 1 galaxies (Fig.~\ref{fig:comp}). The unchanging line profiles (Supplementary Fig. 2), and the 2.6$\sigma$ X-ray source at the onset of the 2015 bump, are also in agreement with Seyfert variability. The 2015 bump is probably not connected to the PS1-10adi event.

\subsection{Radio observations.} 
We retrieved archival data of PS1-10adi targeted by the Karl G. Jansky Very Large Array in D-configuration on 2010 September 10.2 Universal Time, +5.6 d (project code: AS1020). Observations were obtained at 5.0 GHz (C band), and we carried out the data reduction in a standard manner using the Common Astronomy Software Applications package. This yielded an upper limit of $F < 65.252$\,$\upmu$Jy at the location of PS1-10adi, corresponding to a radiated energy of $< 7.8 \times 10^{28}$ erg s$^{-1}$ Hz$^{-1}$, or $< 6.5 \times 10^{28}$ erg s$^{-1}$ Hz$^{-1}$ in the rest frame. However, radio emission of a type IIn supernova with a very massive circumstellar medium could be completely free-free absorbed for up to a few years.

\subsection{Reprocessing shell.} 
Our first-order approximation for a reprocessing shell is a spherical, standing, optically thick (an optical depth $\tau \geq 1$) H-rich shell surrounding a supermassive black hole (SMBH)\cite{loeb97}. The SMBH and shell masses can be estimated $M_{\mathrm{SMBH}} = \sigma_{\mathrm{T}} L_{\mathrm{Edd}} / (4 \pi G m_{p} c)$ and $M_{\mathrm{shell}}  = 8 \pi m_{p} R^{2}_{\mathrm{phot}} \ln(R_{\mathrm{phot}}/R_{\mathrm{shell,in}}) / \sigma_{\mathrm{T}}$, respectively, where $m_{p}$ is the proton mass, $c$ is the velocity of light, $G$ is the gravitational constant, $\sigma_{\mathrm{T}}$ is the Thomson scattering cross-section, $R_{\mathrm{phot}}$ is the photospheric radius and $R_{\mathrm{shell,in}}$ is the inner radius of the shell, and we assume $R_{\mathrm{phot}}/R_{\mathrm{shell,in}} \approx 100$. At light-curve peak, assuming $R_{\mathrm{phot}} \approx R_{\mathrm{BB}} \approx 8 \times 10^{15}$ cm and $L_{\mathrm{Edd}} \approx L_{\mathrm{peak}} \approx 7 \times 10^{44}$ erg s$^{-1}$, where $R_{\mathrm{BB}}$ is the blackbody radius, $L_{\mathrm{Edd}}$ is the Eddington luminosity and $L_{\mathrm{peak}}$ is the peak luminosity, yields $M_{\mathrm{shell}} \approx 9$ $M_{\odot}$ and $M_{\mathrm{SMBH}} \approx 6 \times 10^{6}$ $M_{\odot}$. This is in reasonable agreement with $M_{\mathrm{SMBH}} \approx 10^{7}$ $M_{\odot}$, estimated from the luminosity and FWHM of the H$\alpha$ and H$\beta$ lines\cite{greene05}. An accretion efficiency of $\epsilon = 0.1$ would suggest an accretion rate of $\epsilon c^{2}/L \approx 0.2$ $M_{\odot}$ yr$^{-1}$, with an increase of a factor of $\sim$100 compared with the quiescent level to explain the $\gtrsim$2 mag light-curve brightening. The $R_{\mathrm{BB}}$ at peak is roughly consistent with the size of an AGN broad-line region, and a velocity $v \approx (G M_{\mathrm{SMBH}} / R_{\mathrm{BB}})^{1/2}$ suggests $\sim$3,000$-$4,000 km s$^{-1}$, inconsistent with the observations (900 km s$^{-1}$). It is also probable that $M_{\mathrm{shell}} \ll M_{\mathrm{BLR}}$ in Seyfert galaxies, where $M_{\mathrm{BLR}}$ is the mass of the broad-line region.\cite{baldwin03} The internal energy of the shell is sufficient to cover the radiated energy, and could radiate up to tens of years\cite{loeb97}.

\subsection{Additional candidates.}
PS1-13jw was flagged as a transient in the PS1 Medium Deep survey, field MD03, on 2013 February 6. It was contextually classified as an AGN, and no spectra were obtained. With the aim of discovering PS1-10adi-like events, we recovered PS1-13jw after eyeballing Pan-STARRS Transient Science Server light curves of events with hosts that have been spectroscopically classified as Seyfert galaxies\cite{veron-cetty10} and other AGN\cite{abazajian09} within $z = 0.1 - 0.4$, and have $\geq$20 (in a single filter) automatic detections after reference image subtraction. 

We also searched for PS1-10adi-like events from the Catalina Surveys\cite{drake09} DR2 among the above-mentioned active hosts. Each downloaded light curve was processed by rejecting the brightest and faintest 5\% of the data points. The quiescent level $Q$ was defined as the average of the faintest 50\% of the remaining data. The brightest point of the remaining data was defined as the peak $P$. The standard deviation of the remaining data ($<$20 mag) was defined as the scatter $S$ of the data. Targets with $>$100 data points, $Q < 18.5$ mag, $S < 0.3$ mag, $P - Q > 0.5$ mag and $P - Q > 4 \times S$ were selected for eyeballing. Three potential candidates (Fig.~\ref{fig:comp}) were identified.

\subsection{Host galaxies.}
By +1,216 d, PS1-10adi was roughly the same brightness as the (quiescent) host galaxy. We obtained deep 9$\times$120~s exposures of the field in the R band and I band (seeing $\sim$0.7$''$), and estimate the (point source) 5$\sigma$ image depth as $m_{\mathrm{R}}>23.7$ and $m_{\mathrm{I}}>23.2$~mag. The stacked images reveal neither signs of structure in the host galaxy nor a foreground lensing galaxy (Fig.~\ref{fig:comp}). Furthermore, no lower redshift features are apparent in the spectra. This argues against the possibility that PS1-10adi is amplified by a gravitational lensing event\cite{lawrence16}.

To spatially compare the position of the transient PS1-10adi to that of the host galaxy, we aligned the +182 d ALFOSC R-band follow-up image of PS1-10adi (pixel scale 0.190$''$ per pixel, FWHM $\sim$ 0.6$''$) with the above-mentioned deep +1,216 d StanCam R-band image of the field (pixel scale 0.176$''$ per pixel). A general geometric transformation using the iraf tasks geomap and geotran, based on 36 stars in both images, yielded a root mean square error of 25~mas. The pixel coordinates of PS1-10adi and the centre of the point-like host galaxy were measured using three centering algorithms (Gaussian, centroid and ofilter). This resulted in an average offset of 21~mas (0.12 pixels). We conclude that PS1-10adi is coincident with the host galaxy nucleus within $<$0.046$''$. At an angular size distance of 694~Mpc this corresponds to a projected distance of $<$150~pc from the host nucleus. This is of the same order as the projected $\sim$350 pc host nucleus distance of the superluminous type IIn SN 2006gy\cite{smith07}.

Gaussian components were fitted to the line profiles of the +1,600 d spectrum of the (quiescent) PS1-10adi host galaxy, and the narrow-line component ratios were compared to the theoretical curves\cite{kewley06} in a spectral line ratio Baldwin-Phillips-Terlevich (BPT)\cite{baldwin81} diagram (Supplementary Fig. 4). Aside from the Balmer lines, the spectrum shows weak forbidden lines suggestive of an \HII-dominated host. However, the broad base of the Balmer line profiles suggests the presence of an AGN and supports the classification as a narrow-line Seyfert 1 galaxy. For the host galaxies of other transients in our sample, we used the SDSS DR7 spectra to derive the line ratios. Similar to PS1-10adi, the PS1-13jw and CSS100217 hosts display line ratios that are indicative of ongoing star formation. J094608, and possibly J094806, hosts are potential composite galaxies, whereas the host of J233454 is AGN dominated.

\subsection{Data availability.}
The data that support the findings of this study are available from the corresponding author upon reasonable request.

\end{methods}

\clearpage 

\linespread{1.0}

\newpage
\renewcommand{\tablename}{Supplementary Table}
\setcounter{table}{0}

\begin{table*}
\footnotesize
\caption{PS1-10adi-like transients.}
\centering
\begin{tabular}{cccccccc}
\hline
\hline
Transient & Host & Type\cite{veron-cetty10si} & ROSAT\cite{abazajian09si} & $z$\cite{abazajian09si} & $A_{V, \mathrm{Gal}}$\cite{schlafly11si} & $M_{g, \mathrm{host}}$ & $t_{\mathrm{peak, event}}$\\
 & & & & & (mag) & (mag) & (JD) \\
\hline
PS1-10adi & SDSS J204244.74+153032.1 & - & N & 0.203 & 0.281 & $-19.7$ & 2455443 \\
PS1-13jw & SDSS J084453.56+425744.8 & S1n & Y & 0.345 & 0.086 & $-22.2$ & 2456435 \\
CSS100217 & SDSS J102912.58+404219.7 & S1 & N & 0.147 & 0.040 & $-21.2$ & 2455251 \\
J094806 &  SDSS J094806.55+031801.9 & AGN & N & 0.207 & 0.113 & $-21.5$ & 2454980 \\
J094608 & SDSS J094608.49+351222.4 & S1n & Y & 0.119 & 0.029 & $-20.7$ & 2455735 \\
J233454 & SDSS J233454.07+145712.8 & S1 & N & 0.107 & 0.196 & $-20.5$ & 2455220 \\
\hline
\end{tabular}
\label{table:sample}
\end{table*}

\clearpage

{\footnotesize
\begin{longtable}{ccrcccccc}
\caption{{\normalsize Optical photometry for PS1-10adi (no host subtraction) with the 1$\sigma$ errors given in brackets.}}\\
\hline
\hline
Date & JD & \multicolumn{1}{c}{$t_{\mathrm{rest}}$} & $m_{U}$ & $m_{B}$ & $m_{V}$ & $m_{R}$ & $m_{I}$ & Instrument\\
(UT) & (2400000+) & \multicolumn{1}{c}{(d)} & (mag) & (mag) & (mag) & (mag) & (mag) & \\ 
\hline
\endfirsthead
\caption{continued.}\\
\hline
\hline
Date & JD & \multicolumn{1}{c}{$t_{\mathrm{rest}}$} & $m_{U}$ & $m_{B}$ & $m_{V}$ & $m_{R}$ & $m_{I}$ & Instrument\\
(UT) & (2400000+) & \multicolumn{1}{c}{(d)} & (mag) & (mag) & (mag) & (mag) & (mag) & \\ 
\hline
\endhead
\endfoot
2010-07-14.38 & 55391.88 & $-42$ & - & 18.18(0.03) & 17.73(0.01) & - & - & GPC1 \\
2010-08-15.39 & 55423.89 & $-16$ & - & 17.86(0.01) & 17.41(0.01) & 17.04(0.02) & - & GPC1 \\
2010-08-19.30 & 55427.80 & $-13$ & - & - & - & - & 16.65(0.05) & GPC1 \\
2010-08-29.93 & 55438.43 & $-4$ & - & 17.65(0.01) & 17.39(0.01) & 17.14(0.01) & 16.71(0.01) & ALFOSC \\
2010-09-02.90 & 55442.40 & 0 & 17.08(0.02) & 17.67(0.01) & - & - & - & ALFOSC \\
2010-09-11.07 & 55450.57 & 6 & 17.17(0.02) & 17.75(0.01) & 17.44(0.01) & 17.15(0.01) & 16.73(0.01) & ALFOSC \\
2010-09-17.86 & 55457.36 & 12 & - & 17.77(0.02) & 17.42(0.02) & 17.12(0.01) & 16.70(0.02) & RATCam \\
2010-09-30.88 & 55470.38 & 23 & - & 17.78(0.02) & 17.49(0.01) & 17.16(0.01) & 16.72(0.01) & RATCam \\
2010-10-06.82 & 55476.32 & 28 & 17.29(0.02) & 17.82(0.01) & 17.50(0.01) & 17.22(0.01) & 16.75(0.01) & ALFOSC \\
2010-10-09.83 & 55479.33 & 30 & - & 17.80(0.01) & 17.47(0.01) & 17.16(0.01) & 16.73(0.01) & RATCam \\
2010-10-21.88 & 55491.38 & 40 & 17.27(0.02) & 17.88(0.01) & 17.52(0.01) & 17.24(0.01) & 16.76(0.01) & ALFOSC \\
2010-10-27.81 & 55497.31 & 45 & - & 17.88(0.01) & 17.55(0.01) & 17.22(0.01) & 16.75(0.01) & RATCam \\
2010-11-06.83 & 55507.33 & 53 & 17.36(0.03) & 17.95(0.02) & 17.61(0.01) & 17.28(0.01) & 16.79(0.01) & RATCam \\
2010-11-08.83 & 55509.33 & 55 & 17.45(0.02) & 18.02(0.01) & 17.64(0.01) & 17.33(0.01) & 16.83(0.01) & ALFOSC \\
2010-11-16.91 & 55517.41 & 62 & - & 18.04(0.05) & 17.67(0.02) & - & 16.82(0.02) & RATCam \\
2011-03-28.25 & 55648.75 & 171 & 17.98(0.03) & 18.64(0.02) & 18.13(0.01) & 17.87(0.02) & 17.25(0.01) & ALFOSC \\
2011-04-10.23 & 55661.73 & 182 & - & 18.70(0.02) & 18.23(0.01) & 17.92(0.01) & 17.30(0.01) & ALFOSC \\
2011-06-06.12 & 55718.62 & 229 & 18.28(0.02) & 18.87(0.01) & 18.44(0.01) & 18.15(0.01) & 17.42(0.02) & ALFOSC \\
2011-07-31.97 & 55774.47 & 276 & 18.39(0.04) & 19.09(0.02) & 18.60(0.01) & 18.27(0.01) & 17.57(0.03) & ALFOSC \\
2011-09-04.87 & 55809.37 & 305 & 18.54(0.05) & 19.20(0.02) & 18.77(0.02) & 18.44(0.02) & 17.74(0.01) & ALFOSC \\
2011-10-28.82 & 55863.32 & 349 & 18.60(0.03) & 19.30(0.02) & 18.83(0.01) & 18.52(0.02) & 17.84(0.02) & ALFOSC \\
2012-12-18.80 & 56280.30 & 696 & - & - & 19.49(0.03) & 19.10(0.02) & 18.34(0.03) & StanCam \\
2013-04-26.18 & 56408.68 & 803 & - & - & 19.56(0.06) & 19.19(0.04) & 18.38(0.04) & StanCam \\
2013-09-22.00 & 56557.50 & 926 & 19.50(0.12) & 20.23(0.06) & 19.75(0.04) & 19.29(0.02) & 18.59(0.03) & StanCam \\
2014-09-05.04 & 56905.54 & 1216 & - & - & - & 19.49(0.02) & 18.76(0.02) & StanCam \\
2014-09-31.02 & 56931.52 & 1237 & 19.65(0.05) & 20.41(0.04) & 19.77(0.03) & 19.36(0.03) & 18.59(0.04) & ALFOSC \\
2014-10-06.94 & 56937.44 & 1242 & 19.73(0.08) & 20.40(0.05) & 19.94(0.03) & 19.47(0.02) & 18.70(0.03) & StanCam \\
2015-06-18.19 & 57191.69 & 1454 & 19.48(0.03) & 20.20(0.02) & 19.78(0.02) & 19.39(0.03) & 18.79(0.03) & ALFOSC \\
2015-07-03.07 & 57206.57 & 1466 & 19.46(0.07) & 20.19(0.06) & 19.75(0.04) & 19.37(0.03) & 18.73(0.04) & StanCam \\
2015-08-03.10 & 57237.60 & 1492 & 19.24(0.05) & 19.99(0.04) & 19.50(0.03) & 19.18(0.02) & 18.57(0.03) & StanCam \\
2015-09-03.04 & 57268.54 & 1517 & 19.11(0.05) & 19.85(0.04) & 19.48(0.03) & 19.01(0.03) & 18.47(0.03) & StanCam \\
2015-09-11.89 & 57277.39 & 1525 & - & 19.84(0.02) & 19.47(0.02) & 19.07(0.02) & 18.48(0.02) & IO:O \\
2015-09-27.90 & 57293.40 & 1538 & - & 19.67(0.09) & 19.36(0.06) & 18.97(0.03) & 18.35(0.03) & IO:O \\
2015-10-10.90 & 57306.40 & 1549 & - & 19.88(0.02) & 19.47(0.02) & 19.07(0.01) & 18.42(0.02) & IO:O \\
2015-10-27.92 & 57323.42 & 1563 & - & 20.01(0.09) & 19.49(0.04) & 19.09(0.03) & 18.50(0.03) & IO:O \\
2015-10-27.93 & 57323.43 & 1563 & 19.36(0.09) & 20.01(0.07) & 19.51(0.05) & 19.13(0.03) & 18.56(0.03) & StanCam \\
2015-11-20.86 & 57347.36 & 1583 & - & 20.04(0.05) & 19.53(0.03) & 19.11(0.02) & 18.49(0.03) & IO:O \\
2015-11-23.88 & 57350.38 & 1586 & - & 19.96(0.09) & 19.45(0.04) & 19.01(0.04) & 18.52(0.04) & StanCam \\
2016-03-30.22 & 57477.72 & 1691 & - & 20.00(0.13) & 19.59(0.09) & - & - & IO:O \\
2016-04-13.19 & 57491.69 & 1703 & - & 20.03(0.06) & 19.60(0.03) & 19.10(0.02) & 18.49(0.03) & IO:O \\
2016-04-27.22 & 57505.72 & 1715 & - & 20.18(0.07) & 19.65(0.04) & 19.20(0.02) & 18.55(0.03) & IO:O \\
2016-05-25.18 & 57533.68 & 1738 & - & 20.21(0.06) & 19.70(0.04) & 19.28(0.02) & 18.63(0.03) & IO:O \\
2016-07-06.10 & 57575.60 & 1773 & - & 20.14(0.03) & 19.67(0.02) & 19.23(0.02) & 18.58(0.02) & IO:O \\
2016-08-03.02 & 57603.52 & 1796 & - & 20.20(0.03) & 19.76(0.02) & 19.26(0.02) & 18.66(0.03) & IO:O \\
2017-03-15.26 & 57827.76 & 1982 & - & 20.02(0.11) & 19.60(0.07) & 19.12(0.04) & 18.56(0.04) & IO:O \\
2017-04-08.24 & 57851.74 & 2002 & - & 20.26(0.07) & 19.61(0.05) & 19.21(0.02) & 18.66(0.04) & IO:O \\
\hline
2017-04-24.21 & 57867.71 & 2016 & - & 20.12(0.05) & 19.70(0.03) & 19.27(0.02) & 18.65(0.02) & IO:O \\
2017-05-13.19 & 57886.69 & 2031 & - & 20.24(0.08) & 19.65(0.04) & 19.26(0.02) & 18.64(0.03) & IO:O \\
2017-05-28.07 & 57901.57 & 2044 & - & 20.25(0.05) & 19.60(0.03) & 19.18(0.02) & 18.58(0.03) & IO:O \\
2017-06-11.09 & 57915.59 & 2055 & - & 20.33(0.11) & 19.78(0.07) & 19.19(0.03) & 18.58(0.04) & IO:O \\
2017-06-25.11 & 57929.61 & 2067 & - & 20.22(0.15) & 19.74(0.60) & 19.36(0.08) & 18.72(0.06) & IO:O \\
2017-07-10.08 & 57944.58 & 2079 & - & 20.41(0.11) & 19.64(0.06) & 19.38(0.03) & 18.74(0.05) & IO:O \\
2017-08-01.06 & 57966.56 & 2098 & - & 20.28(0.03) & 19.79(0.02) & 19.33(0.02) & 18.70(0.03) & IO:O \\
\hline
\label{table:phot}
\end{longtable}}

\clearpage

\begin{table*}
\footnotesize
\caption{IR photometry for PS1-10adi (no host subtraction) with the 1$\sigma$ errors given in brackets.}
\centering
\begin{tabular}{ccrcccccc}
\hline
\hline
Date & JD & \multicolumn{1}{c}{$t_{\mathrm{rest}}$} & $m_{J}$ & $m_{H}$ & $m_{K}$ & $m_{\mathrm{3.4 \mu m}}$ & $m_{\mathrm{4.6 \mu m}}$ & Instrument\\
(UT) & (2400000+) & \multicolumn{1}{c}{(d)} & (mag) & (mag) & (mag) & (mag) & (mag) & \\ 
\hline
2010-05-09.90 & 55326.40 & $-97$ & - & - & - & 14.48(0.02) & 13.67(0.03) & WISE \\
2010-09-29.88 & 55469.38 & 22 & 16.06(0.02) & 15.39(0.02) & 14.61(0.02) & - & - & NOTCam \\
2010-10-28.87 & 55498.37 & 46 & 15.95(0.03) & 15.11(0.04) & 14.39(0.02) & - & - & NOTCam \\
2010-11-05.20 & 55505.70 & 52 & - & - & - & 13.29(0.01) & 12.68(0.01) & WISE \\
2010-11-18.85 & 55519.35 & 63 & 15.94(0.03) & 15.17(0.03) & 14.32(0.02) & - & - & NOTCam \\
2010-12-12.84 & 55543.34 & 83 & 15.93(0.03) & 15.14(0.02) & 14.24(0.02) & - & - & NOTCam \\
2011-05-16.20 & 55697.70 & 212 & 16.33(0.02) & 15.41(0.01) & 14.41(0.02) & - & - & NOTCam \\
2011-06-11.11 & 55723.61 & 233 & 16.44(0.02) & 15.50(0.02) & 14.41(0.02) & - & - & NOTCam \\
2011-08-14.14 & 55787.64 & 286 & 16.62(0.02) & 15.67(0.02) & 14.55(0.02) & - & - & NOTCam \\
2011-09-14.03 & 55818.53 & 312 & 16.71(0.02) & 15.75(0.02) & 14.63(0.02) & - & - & NOTCam \\
2012-05-07.18 & 56054.68 & 508 & 17.20(0.02) & 16.26(0.02) & 15.07(0.02) & - & - & NOTCam \\
2012-07-05.13 & 56113.63 & 557 & 17.24(0.02) & 16.34(0.03) & 15.14(0.01) & - & - & NOTCam \\
2012-08-06.15 & 56145.65 & 584 & 17.39(0.02) & 16.48(0.02) & 15.28(0.02) & - & - & NOTCam \\
2012-09-07.96 & 56178.46 & 611 & 17.46(0.02) & 16.53(0.02) & 15.34(0.03) & - & - & NOTCam \\
2013-10-19.95 & 56585.45 & 950 & 17.90(0.03) & 17.06(0.03) & 15.89(0.02) & - & - & NOTCam \\
2013-12-16.82 & 56643.32 & 998 & 17.79(0.03) & 17.03(0.02) & 15.92(0.03) & - & - & NOTCam \\
2014-05-11.18 & 56788.68 & 1119 & - & - & - & 14.32(0.02) & 13.27(0.02) & WISE \\
2014-05-20.20 & 56797.70 & 1126 & 17.92(0.03) & 17.21(0.03) & 16.10(0.03) & - & - & NOTCam \\
2014-06-10.19 & 56818.69 & 1144 & 17.95(0.03) & 17.17(0.03) & 16.10(0.03) & - & - & NOTCam \\
2014-07-10.14 & 56848.64 & 1168 & 17.97(0.03) & 17.19(0.03) & 16.13(0.02) & - & - & NOTCam \\
2014-08-07.11 & 56876.61 & 1192 & 17.97(0.04) & 17.27(0.05) & 16.20(0.04) & - & - & NOTCam \\
2014-09-05.09 & 56905.59 & 1216 & 17.97(0.03) & 17.23(0.03) & 16.14(0.03) & - & - & NOTCam \\
2014-10-06.97 & 56937.47 & 1242 & 17.98(0.02) & 17.26(0.03) & 16.22(0.02) & - & - & NOTCam \\
2014-11-08.82 & 56970.32 & 1270 & - & - & - & 14.46(0.02) & 13.45(0.03) & WISE \\
2015-07-03.11 & 57206.61 & 1466 & 17.89(0.03) & 17.19(0.03) & 16.26(0.03) & - & - & NOTCam \\
2015-08-03.08 & 57237.58 & 1492 & 17.76(0.03) & 17.08(0.03) & 16.20(0.02) & - & - & NOTCam \\
2015-09-03.01 & 57268.51 & 1517 & 17.73(0.02) & 17.00(0.04) & 16.22(0.03) & - & - & NOTCam \\
2015-10-27.96 & 57323.46 & 1563 & 17.67(0.03) & 17.08(0.02) & 16.17(0.02) & - & - & NOTCam \\
2015-11-23.85 & 57350.35 & 1585 & 17.71(0.03) & 17.08(0.05) & 16.14(0.02) & - & - & NOTCam \\
2015-12-22.82 & 57379.32 & 1610 & 17.86(0.29) & 17.29(0.31) & 16.01(0.12) & - & - & NOTCam \\
2016-03-25.22 & 57472.72 & 1687 & 17.76(0.03) & 17.13(0.04) & 16.19(0.04) & - & - & NOTCam \\	
\hline
\end{tabular}
\label{table:phot_NIR}
\end{table*}

\clearpage

\begin{table*}
\footnotesize
\caption{\textit{Swift} UVOT (Vega system) UV photometry ($\lambda_{UVW2} = 1928$~\AA, $\lambda_{UVM2} = 2246$~\AA, and $\lambda_{UVW1} = 2600$~\AA) and XRT X-ray observations (0.3$-$10 keV) for PS1-10adi (no host subtraction) with the 1$\sigma$ errors given in brackets.}
\centering
\begin{tabular}{ccrcccccc}
\hline
\hline
Date & JD & \multicolumn{1}{c}{$t_{\mathrm{rest}}$} & $m_{UVW2}$ & $m_{UVM2}$ & $m_{UVW1}$ & counts$_{\mathrm{X}}$ & $L_{\mathrm{X}}$ & $t_{\mathrm{X}}$ \\
(UT) & (2400000+) & \multicolumn{1}{c}{(d)} & (mag) & (mag) & (mag) & ($10^{-3}$~s$^{-1}$) & ($10^{43}$~erg~s$^{-1}$) & (s) \\ 
\hline
2010-10-06.8  & 55476.3 & 28 & 17.75(0.05) & 17.57(0.05) & 17.25(0.05) & $<$2.1 & $<$1.1 & 5224 \\
2010-10-07.8  & 55477.3 & 29 & - & - & - & $<$2.6 & $<$1.3 & 4211 \\
2010-10-09.9  & 55479.4 & 30 & - & 17.47(0.07) & - & $<$12 & $<$6.2 & 979 \\
2010-12-08.9  & 55539.4 & 80 & 17.98(0.06) & 17.86(0.07) & 17.56(0.06) & $<$2.4 & $<$1.2 & 4737 \\
2015-07-18.9  & 57222.4 & 1479 & 19.54(0.09) & - & - & $<$5.1 & $<$2.6 & 2242 \\
2015-07-19.9  & 57223.4 & 1480 & - & 19.37(0.09) & - & $3.5(1.4)$ & $1.8(0.7)$ & 3424 \\
2015-07-20.5  & 57224.0 & 1480 & - & - & 19.22(0.10) & $3.1(2.2)$ & $1.6(1.1)$ & 1314 \\
\hline
\end{tabular}
\label{table:phot_UV_X-ray}
\end{table*}

\clearpage

\setcounter{figure}{0} 
\renewcommand\figurename{\textbf{Supplementary Figure}}
\renewcommand{\thefigure}{\textbf{\arabic{figure}}}

\begin{figure}
 \centering
 \includegraphics[width=0.8\linewidth]{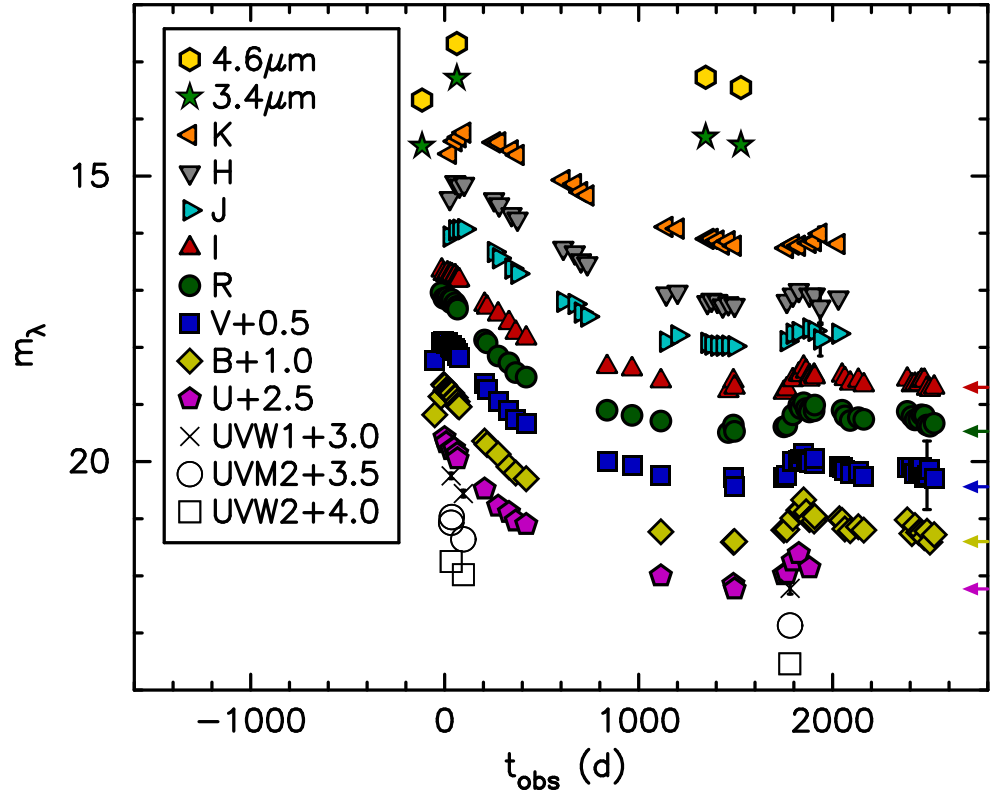}
 \caption{\textbf{Observed multi-colour light curves of PS1-10adi and its host galaxy.} Arrows on the right side of the figure show the pre-discovery \textit{UBVRI} magnitudes of the (quiescent) host galaxy converted\cite{jester05si} from the SDSS DR7 \textit{ugri} magnitudes. The light curves of PS1-10adi (no host subtraction) appear to have plateaued by day $\sim$1200 to 1500 to this level. The different symbols show different filters, and some of the light curves are shifted for clarity as indicated in the legend. The 1$\sigma$ uncertainties are typically smaller than the symbols.}
 \label{fig:phot}
\end{figure}

\begin{figure}
 \centering
 \includegraphics[width=0.8\linewidth]{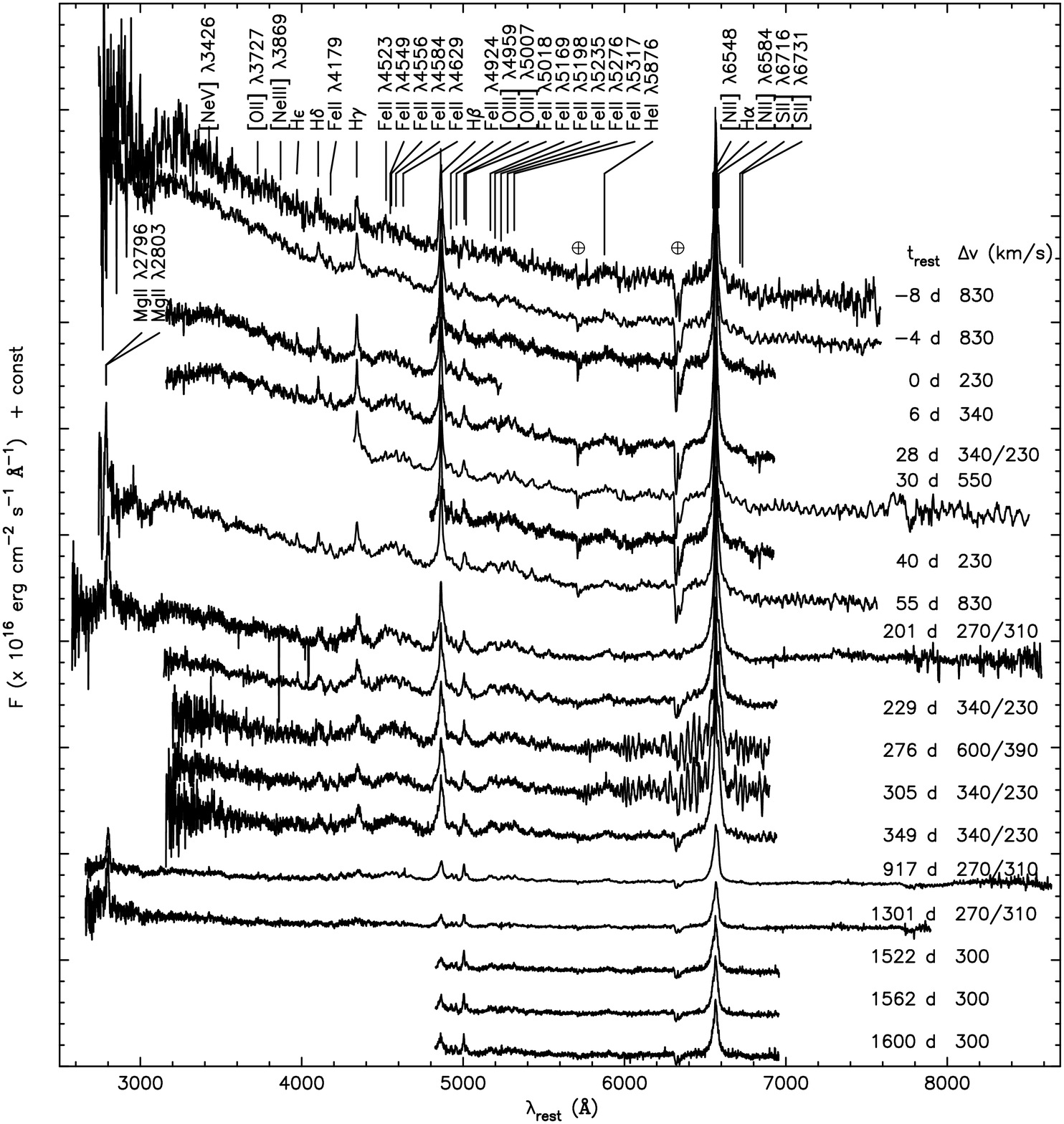}
 \caption{\textbf{Spectroscopic time series of PS1-10adi.} The spectra have been corrected for  Galactic extinction, shifted to the rest frame, and offset vertically for clarity. The wavelengths of the strongest telluric features are indicated by $\oplus$ symbols. Epochs are labelled respective to the estimated optical peak, in the rest frame, and the spectral resolutions are indicated.}
 \label{fig:spectra}
\end{figure}

\begin{figure}
 \centering
 \includegraphics[width=0.8\linewidth]{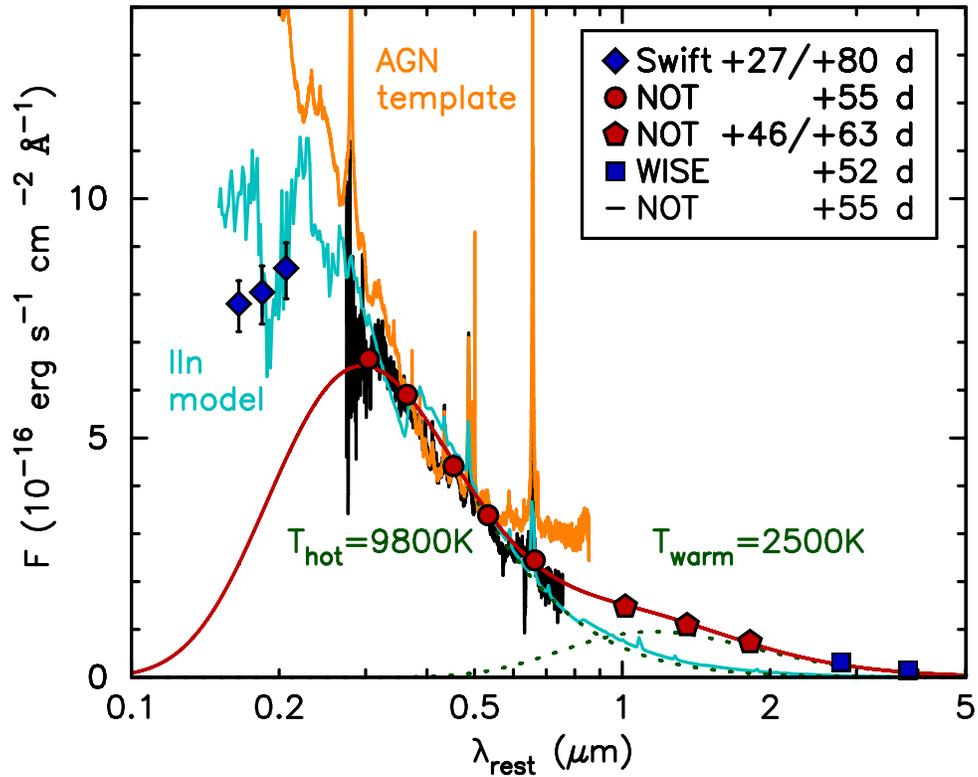}
 \caption{\textbf{Spectral energy distribution (SED) of PS1-10adi event around +55 d.} A single blackbody cannot be used to accurately match the \textit{UBVRIJHK} photometry; however, a two-component fit (individual components displayed with dotted lines) well describes the SED from the optical to near-infrared. An AGN template spectrum\cite{vandenberk01si} (orange) is shown for comparison. Also shown is a synthetic spectrum (model X@41.7d, binned for clarity) of an interacting type IIn supernova\cite{dessart15si} (cyan). Both AGN and supernova spectra have been scaled to match the optical SED of PS1-10adi. The 1$\sigma$ uncertainties are typically smaller than the symbols.}
 \label{fig:SED}
\end{figure}

\begin{figure}
 \centering
 \includegraphics[width=0.8\linewidth]{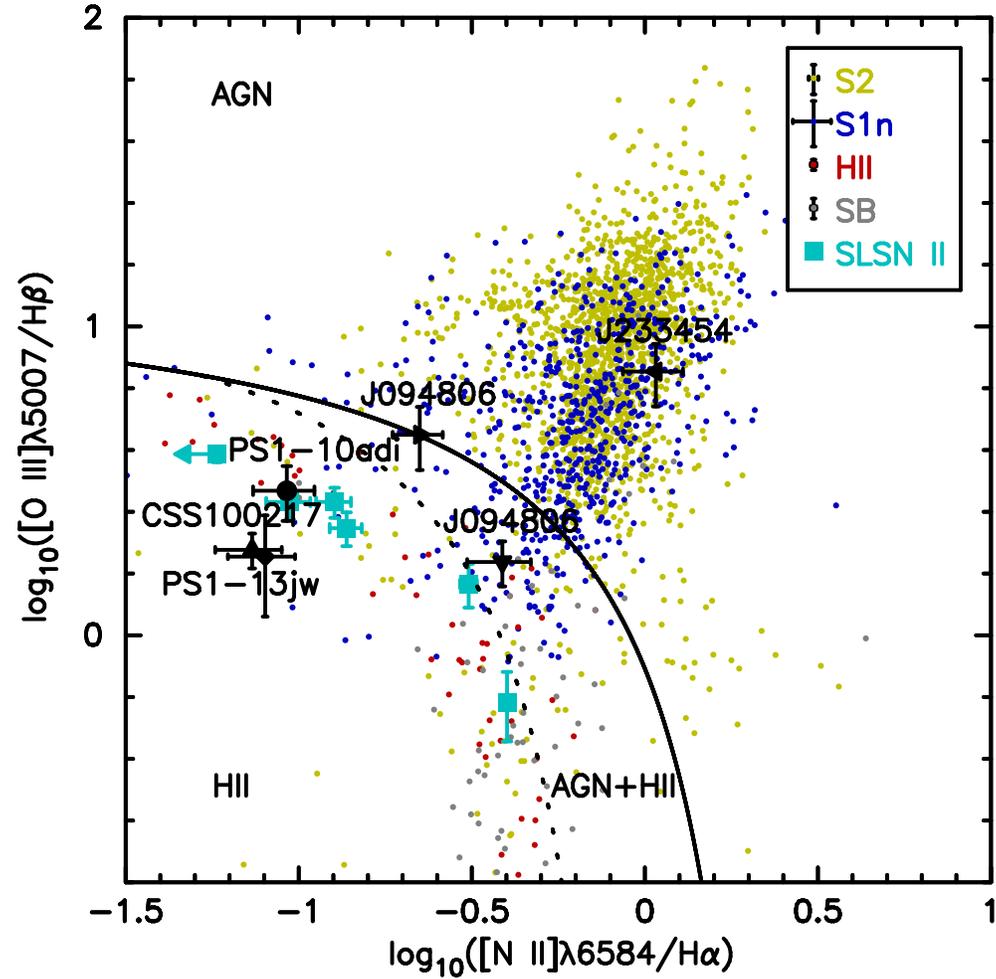}
 \caption{\textbf{PS1-10adi-like transients in a [\NII]$\lambda$6584/H$\alpha$ vs. [\OIII]$\lambda$5007/H$\beta$ BPT diagram.} PS1-10adi is shown with a black circle, and similar events with other black symbols. The empirical classification curves\cite{kewley06si} separating the AGN and the star-forming galaxies are shown. Composite galaxies are expected to fall in the area between the solid line and the dotted line. The diagram includes galaxies with Seyfert 2 and \HII\ dominated nuclei\cite{veron-cetty10si}, as well as highly starbursting (SB) galaxies\cite{sanders03si} with $\log L_{\mathrm{IR}}>10.0$ $L_{\odot}$, with line ratios from SDSS DR7. Furthermore, a sample of narrow-line Seyfert 1 galaxies\cite{zhou06si} and host galaxies of type IIn superluminous supernovae\cite{leloudas15si} are also included. The diagram refers to line ratios of the narrow-line components. Representative errors for Seyfert, \HII\ and SB dominated galaxies are shown in the legend.}
 \label{fig:BPT}
\end{figure}

\clearpage

\subsection{Supplementary Information References}

\end{document}